\documentclass[twocolumn, linenumbers]{aastex62}

\usepackage{graphicx}
\usepackage{url}
\usepackage{amsmath}

\usepackage{tabularx,diagbox,ragged2e}
\usepackage{float}
\usepackage{aliascnt}
\newaliascnt{eqfloat}{equation}
\newfloat{eqfloat}{h}{eqflts}
\floatname{eqfloat}{Equation}

\newcommand*{\ORGeqfloat}{}
\let\ORGeqfloat\eqfloat
\def\eqfloat{%
  \let\ORIGINALcaption\caption
  \def\caption{%
    \addtocounter{equation}{-1}%
    \ORIGINALcaption
  }%
  \ORGeqfloat
}

\begin{document}

    \title{Analysis of a Century's Worth of AR Scorpii Photometry from DASCH and ASAS-SN}
    \author[0000-0001-8596-4746]{Erik Peterson}
    \affiliation{Department of Physics, University of Notre Dame, Notre Dame, IN 46556 USA}
    \author[0000-0001-7746-5795]{Colin Littlefield}
    \affiliation{Department of Physics, University of Notre Dame, Notre Dame, IN 46556 USA}
    \author[0000-0003-4069-2817]{Peter Garnavich}
    \affiliation{Department of Physics, University of Notre Dame, Notre Dame, IN 46556 USA}

\received{June 7, 2019}
\accepted{June 17, 2019}
\published{September 2, 2019}
    
\begin{abstract}
AR Scorpii (AR Sco) is a binary star system containing the only known white dwarf (WD) pulsar. Previously reported photometric datasets only provide coverage back to 2005, but we extend the observational baseline for AR~Sco back to the beginning of the 20th century by analyzing observations from the Digital Access to a Sky Century at Harvard project (DASCH). We find that the orbital waveform of AR Sco remained constant across that baseline with no significant deviations from its average brightness. This result strongly suggests that the absence of accretion in modern observations is a long-term feature of AR Sco. Additionally, the DASCH light curve provides an opportunity to test an earlier prediction that an obliquity of the WD would result in a precessional period observable in long-term studies of the orbital light curve. The DASCH observations do not indicate the presence of such a period, and we show that previous, inconclusive tests of this hypothesis were insensitive to the existence of a precessional period. Furthermore, the long DASCH baseline enables us to constrain the rate of change of the orbital frequency to $\dot{\nu} \lesssim 3.8\times10^{-20}$ Hz s$^{-1}$, constraining the efficacy of magnetic braking as a mechanism of angular-momentum loss in this system. Finally, we discuss how the combination of the orbital waveform's stability, high amplitude, and short period should make it possible to identify additional WD pulsars in all-sky survey data.

\end{abstract}


\section{Introduction}
    \label{sec:intro}
    AR Sco is a binary star system with a white dwarf (WD) and an M-dwarf companion that orbits once every 3.56 hr. It generates highly periodic electromagnetic pulses every 1.97 minutes, during which it brightens by as much as a factor of 4. These pulses consist of synchrotron radiation from relativistic electrons accelerated by the interaction between the WD's magnetosphere and the companion \citep{marsh, buckley}. While the beat pulses (the difference between the WD spin frequency and orbital frequency) were initially hypothesized to arise on the companion star, a polarimetric study by \citet{pb2} challenged this inference by showing that their origin appears to be fixed in the rotational frame of the WD, possibly within the WD's magnetosphere. \citet{garnavich} presented time-series spectroscopy that revealed the presence of slingshot prominences on the companion star and suggested that the beat pulses result from rapid magnetic reconnection events, essentially stellar flares, induced by the magnetic interaction of the two magnetospheres. 
    
     \citet{buckley} described the system as the only known WD pulsar on the grounds that its rapid pulses are powered by the spindown of the rapidly rotating, highly magnetized WD---though the mechanism of emission is different than in neutron-star pulsars. The existence of the WD's spindown was briefly called in doubt when \citet{pb1} showed that the spin-frequency derivative from the \citet{marsh} discovery paper did not agree with subsequent observations. However, \citet{stiller} resolved this controversy by measuring a spin-frequency derivative of $\dot{\nu}_{spin} =  (−5.14\  \pm \ 0.32)\    10^{-17}\  \textrm{Hz} \  \textrm{s}^{-1}$ and calculating that the resulting spindown energy was large enough to power AR Sco's characteristic pulses.
    
    
    While AR Sco is perhaps most famous for its rapid synchrotron pulses, its orbital light curve is notable in its own right. The orbital modulation has a remarkably large peak-to-peak amplitude of $\sim1.8$~mag in the $V$ band \citep{littlefield}, and maximum light occurs near orbital phase $\sim$0.4 \citep{marsh, littlefield}, where orbital phase 0.0 is defined as the secondary's time of inferior conjunction. Exploring this peculiarity, \citet{katz} advances two possible explanations. In his first, he proposes that a magnetic bow wave forms on the trailing hemisphere of the companion star due to enhanced magnetic dissipation caused by the rapid rotation of the WD's magnetosphere. If the rotation of the WD is prograde, the asymmetric heating of the secondary would cause the orbital maximum to precede superior conjunction, as observed.   Alternatively, \citet{katz} suggests that the WD could be an oblique rotator whose spin axis is misaligned with respect to the orbital axis, resulting in enhanced magnetic dissipation twice every orbit when the companion crosses the plane swept out by the WD's magnetic axis. For brevity, we hereafter refer to this plane as the WD magnetic plane. \citet{katz} identified a specific observational prediction for this model: a WD precessional period of \mbox{20-200 yr.} \citet{littlefield} attempted to test the precessional hypothesis by using photometry with a baseline of 11 yr with indeterminate results, but as we will show, their methodology was insensitive to the presence of a precessional period.
    
    \citet{garnavich} further explored the orbital light curve by showing that two basic models were each capable of reproducing the orbital waveform, though neither was unique. The first model required the presence of two localized hotspots on the companion star: one near the L1 region and a second on the hemisphere facing away from the WD. Their second model invoked the \citet{katz} oblique-rotator hypothesis and showed that geometric viewing-angle effects could reproduce the observed orbital light curve if the secondary experienced enhanced heating twice each orbit when it crosses through the WD magnetic plane.
    
    
    

    The existing literature has not explored AR Sco's behavior prior to 2005 \citep{marsh, littlefield}, so its behavior during the 20th~century is largely unknown. To fill this void, we use digitized photographic plates to study AR Sco's orbital light curve across a century-long baseline.

\section{Data }

    \subsection{DASCH}
\begin{figure}[!hbt]
		\begin{center}
		\includegraphics[width=\columnwidth]{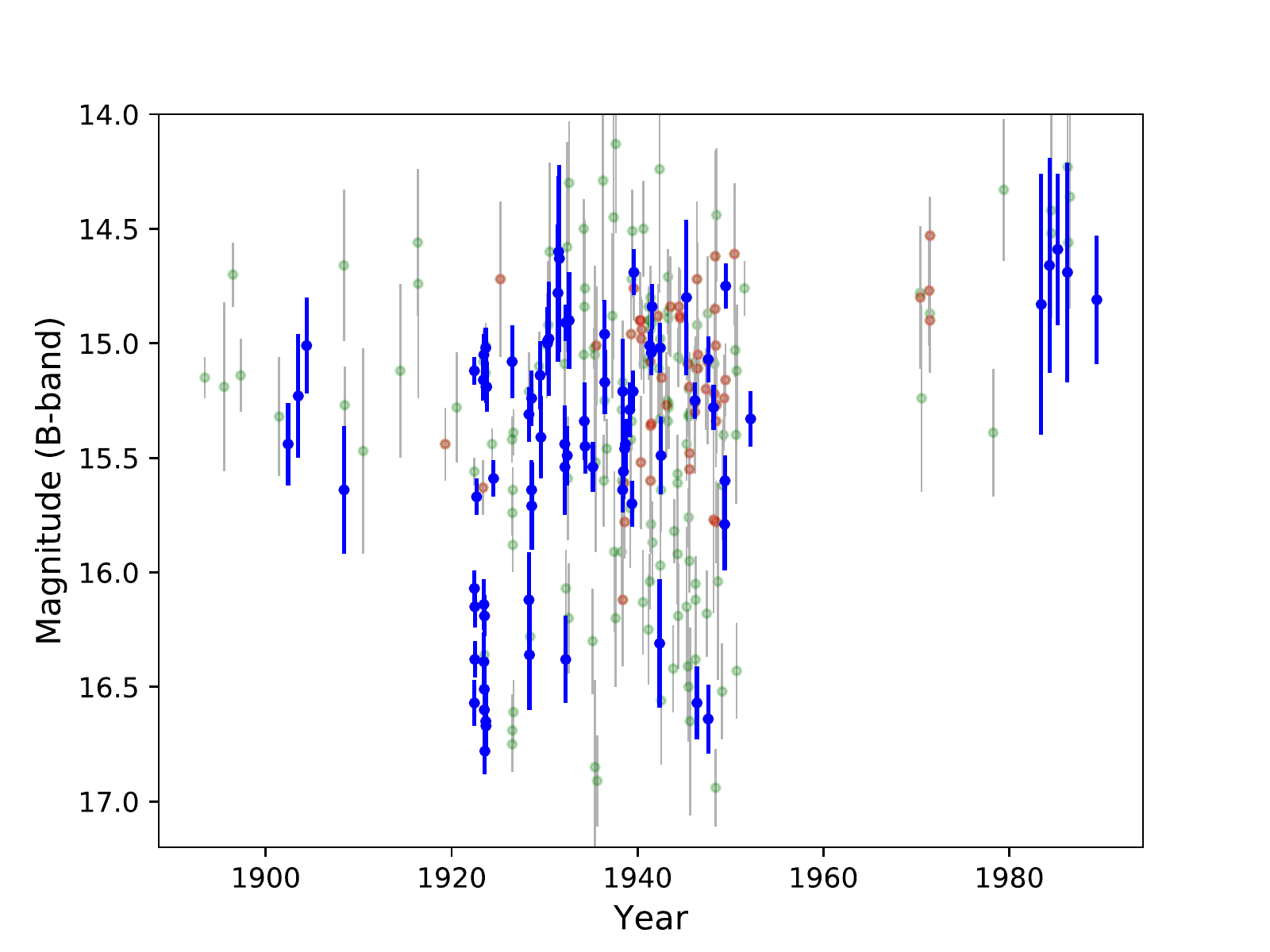}
		\caption{The DASCH light curve. We filtered out the green points according to our quality flag criteria, and the red points were filtered out for exposure times greater than 75 minutes. The blue points represent our data after applying the quality cuts described in the text. Several discrepant points are not shown because either their pipeline uncertainties exceeded 1 mag or because an obvious plate defect was present at their position. The astrophysical variability in this figure is attributable to the orbital modulation because even the shortest exposure times spanned many beat pulses. We note no prolonged bright or dim states for AR Sco over the course of this baseline.}
		\label{fig:DASCH_year_vs_mag}
		\end{center}
	\end{figure}

     The Digital Access to a Sky Century at Harvard project (DASCH) is an ongoing effort to digitize 100 years of sky surveys from photographic plates  \citep{DASCH}. We downloaded the light curve of AR Sco from the DASCH pipeline using the default search radius of 5'' and the APASS $B$ photometric calibration, which, according to the DASCH documentation,\footnote{\url{http://dasch.rc.fas.harvard.edu/lightcurve.php}} is the magnitude calibration that yields the most accurate photometry. With this dataset, our baseline runs back to 1902, albeit with occasional gaps, the most conspicuous of which is the Menzel gap in the 1950s and 1960s (Fig. \ref{fig:DASCH_year_vs_mag}).

    The plates contained within DASCH vary widely in quality, so an important part of using DASCH data is filtering out low-quality observations. To that end, each DASCH observation is accompanied by a series of flags containing information about the reliability of the measurement.\footnote{A full explanation of the quality flags is available from the DASCH website at \url{http://dasch.rc.fas.harvard.edu/database.php}. } After extensive experimentation with various quality criteria, we arrived at the following data cuts. We exclude any measurements with AFLAGS with bit positions of 6-12, 14, or 30; BFLAGS with bit positions of 1, 27, or 29; and quality flags of 0-6, 11, and 12. We also excluded plates whose limiting magnitude was brighter than 14.5, since these plates lack the requisite depth to detect AR Sco. 
    
    Given our objective of studying AR Sco's short-period variability across the DASCH baseline, it is important to know the exposure time of each plate. This information is not included in the datafile downloaded from the DASCH pipeline, so we determined the exposure times by retrieving metadata for each plate from the DASCH website. Of the 127 plates that withstood our earlier quality cuts, 77 had exposures times of 75 minutes or less. The shortest exposures were just 10 minutes, while at the other extreme, one plate had an exposure time of 4~hr. Another 41 plates had exposure times of 2~hr. Since even the shortest exposures averaged across five beat cycles, the DASCH plates cannot provide insight into the behavior of the fast pulses. They do, however, probe the comparatively slow orbital modulation.
    
    A subtle effect of the vastly different exposure times is that long exposures will smear the orbital light curve, distorting its shape and diluting the amplitude of variability. For some science objectives, such as checking for a cessation of the orbital variability or a major brightening event, this effect does not matter. However, it is an important consideration when comparing the morphology of the orbital waveform at different epochs across the DASCH baseline. We simulated this effect by taking a Fourier-series representation of AR Sco's orbital waveform from the \citet{stiller} photometry and convolving it to different exposure times. We found that the shape of the waveform began to change quickly for exposure times longer than 75 minutes, although this is a subjective determination. Thus, in our timing analysis, we analyzed only the plates with exposure times of 75 minutes or shorter.
    
    
    After applying these cuts, we confirmed that the remaining plates were all blue sensitive by ensuring that the plate class for each plate was L (L.~Smith 2019, private communication). Therefore, no red- or yellow-sensitive plates were included in our analysis.

    \begin{deluxetable}{ccc}
    \caption{DASCH identifiers, APASS $V$ magnitudes, and APASS $B-V$ colors for the 20 check stars.    \label{table:checkstars}}
    \tablehead{\colhead{DASCH ID} & \colhead{$V$} & \colhead{$B-V$}}
    \startdata
    APASS J161311.6-245018 & 14.486 & 0.394 \\
    APASS J161316.8-242651 & 14.820 & 0.357 \\
    APASS J161317.0-211140 & 14.387 & 0.213 \\
    APASS J161330.0-240124 & 14.742 & 0.462 \\
    APASS J161446.4-234052 & 14.570 & 0.432 \\
    APASS J161458.0-211527 & 14.659 & 0.371 \\
    APASS J161508.1-214510 & 14.457 & 0.443 \\
    APASS J161513.7-215027 & 14.586 & 0.434 \\
    APASS J161604.2-240736 & 14.206 & 0.342 \\
    APASS J161635.4-220350 & 14.569 & 0.389 \\
    APASS J161638.8-230840 & 14.931 & 0.092 \\
    APASS J161706.8-212438 & 14.536 & 0.487 \\
    APASS J161816.8-230108 & 14.648 & 0.431 \\
    APASS J161834.8-230036 & 14.838 & 0.432 \\
    APASS J161923.0-223234 & 14.574 & -0.070 \\
    APASS J161924.9-225535 & 14.352 & 0.464 \\
    APASS J161940.5-224759 & 14.875 & 0.412 \\
    APASS J162105.6-221803 & 14.717 & 0.163 \\
    APASS J162735.4-213954 & 14.598 & 0.354 \\
    APASS J162933.3-210811 & 14.835 & 0.350
    \enddata
    \end{deluxetable}
    
    Furthermore, in an effort to better understand any systematic errors in the DASCH plates for AR Sco, we selected 20 check stars whose DASCH identifiers and photometric properties are listed in Table~\ref{table:checkstars}. We selected these stars by using the APASS catalog to identify stars within a radius of 4$^{\circ}$ of AR Sco that satisfied two additional criteria: $14<V<15$ and $B-V<0.5$. These ranges approximate AR Sco's color \citep{garnavich} and brightness \citep{littlefield} near its orbital maximum. Ideally, the check stars would be closer to AR Sco, but since AR Sco lies along a heavily reddened line of sight, there are relatively few nearby blue stars.  We extracted their light curves using the DASCH pipeline and applied the same quality cuts as we did with AR Sco. We also inspected images of the check stars before and after the Menzel gap to ensure that the stars did not suffer from blending.\footnote{As \citet{schaefer} has pointed out, the coarser plate scale of the post-Menzel-gap plates creates an increased risk of blending. Blended stars would appear to brighten in post-gap data, creating the misleading appearance of a long-term brightening.} In Sec.~\ref{sec:checkstars}, we use these check stars to identify a systematic trend in the post-Menzel-gap plates of AR Sco.
    

	\subsection{ASAS-SN}
	In order to extend the baseline in more recent years, we incorporated all $V$-band photometry from the All-Sky Automated Survey for Supernovae \citep[ASAS-SN;][]{shappee, kochanek}. While the DASCH light curve for AR Sco has not been analyzed previously, \citet{littlefield} previously analyzed ASAS-SN observations of AR Sco from 2014, 2015, and 2016. We add to this baseline ASAS-SN observations from 2017 and 2018. Compared to the Harvard plates, the ASAS-SN data are much more homogeneous and tend to be of higher quality, providing a useful point of comparison for the DASCH light curve.
	
	In total, our dataset runs from 1902 until 2018, albeit with large gaps in the middle of the 20th century, in the 1990s, and in the early 2000s. 

	\section{Analysis}

	\subsection{Long-term Photometric Stability}
	
	In the century-long light curve (Fig. \ref{fig:DASCH_year_vs_mag}), we note no prolonged bright or dim states. While there are many more detections of AR Sco fainter than 16th magnitude between 1920 and 1950, this is simply because the limiting magnitudes of those plates were deeper than at the beginning and end of the DASCH light curve---a consequence of the many different telescopes, emulsions, and observing sites that collectively make up the DASCH light curve.
	
\begin{figure}[hb]
		\begin{center}
		\includegraphics[width=\columnwidth]{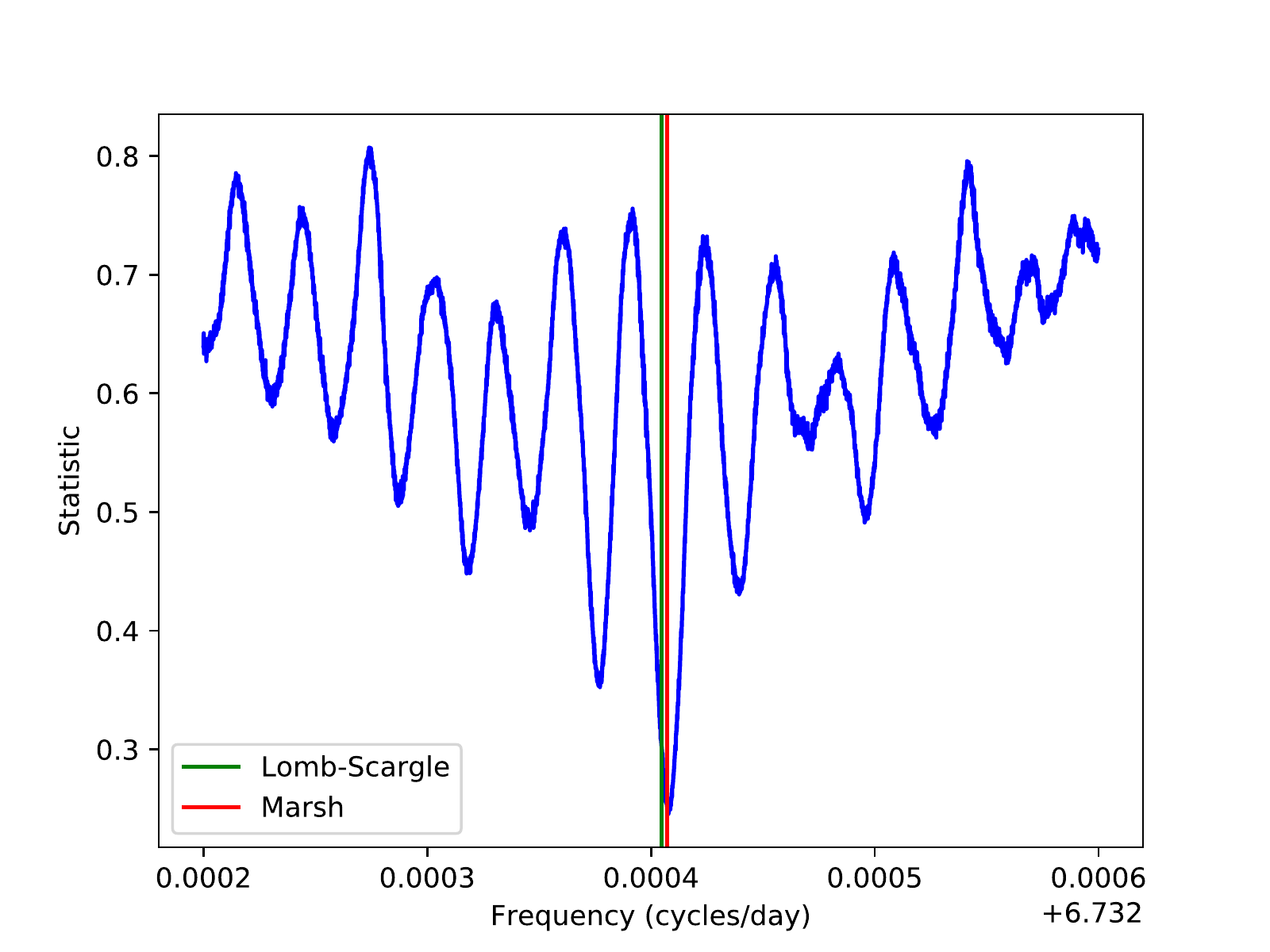}
		\caption{Period analysis using phase-dispersion minimization. The orbital periods from \citet{marsh} and from a Lomb-Scargle analysis are plotted for reference. We take the minimum outputted statistic to be the true orbital frequency for AR Sco. We note that the value obtained from our period analysis method is in strong agreement with the orbital period obtained by \citet{marsh}.}
		\label{fig:pdm_phase}
		\end{center}
		
	\end{figure}

	As shown in Fig.~\ref{fig:pdm_phase}, we measured the orbital period in the combined DASCH and ASAS-SN dataset to be 0.14853526(5) days by using phase-dispersion minimization \citep[PDM;][]{PDM} after first applying a zero-point offset to better align the $V$-band ASAS-SN data with the blue-sensitive DASCH measurements. In order to ensure that the sheer number of the ASAS-SN measurements did not overpower the data from DASCH in the PDM analysis, we incorporated a random selection of only 20\% of the data from ASAS-SN, which we found yielded a reasonable density of ASAS-SN points relative to DASCH observations.

	\begin{figure}[!hbt]
		\begin{center}
		\includegraphics[width=\columnwidth]{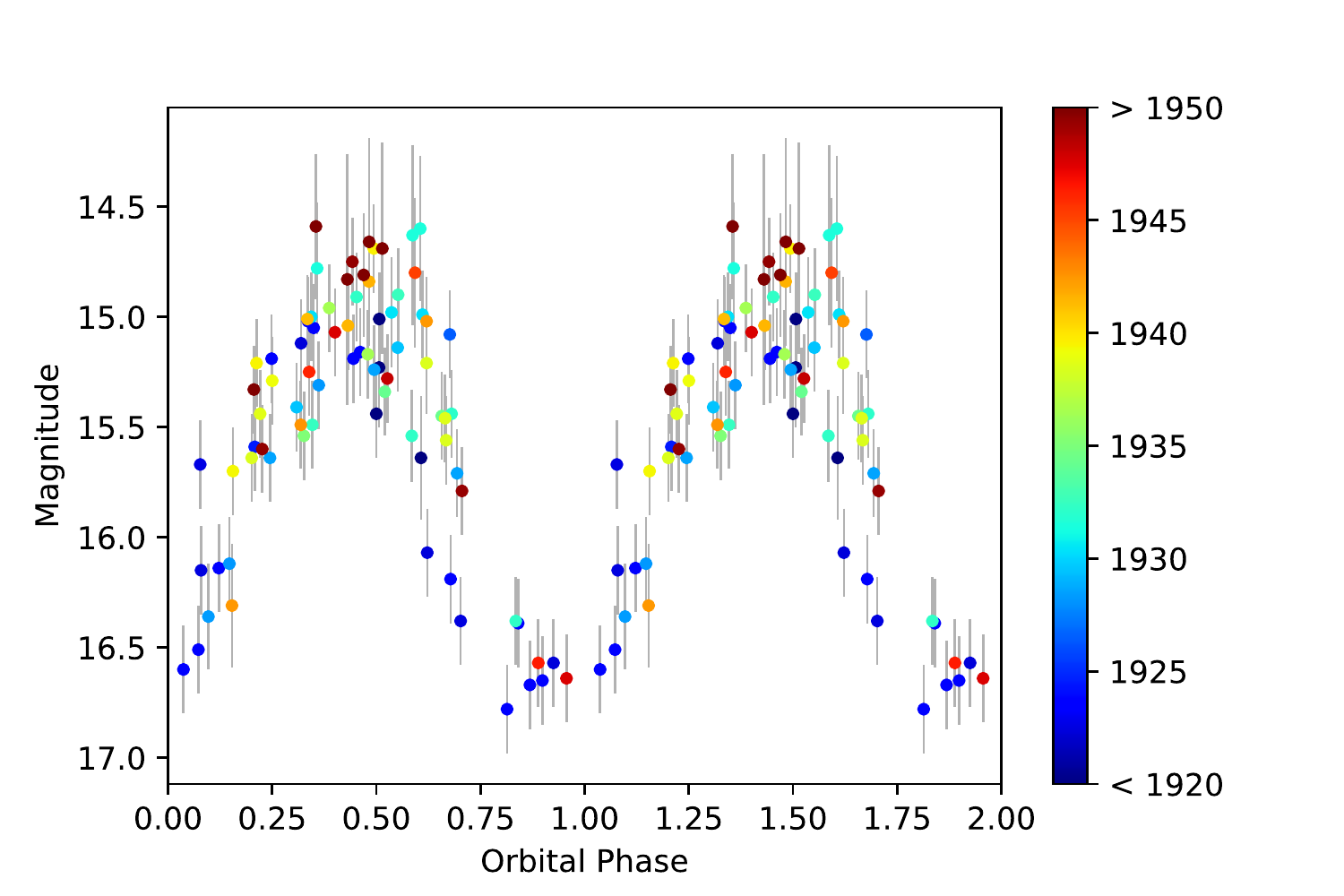}
		\caption{Phased light curve of the DASCH data, with color corresponding to the year of data collection.}
		\label{fig:DASCH_phased_pdm}
		\end{center}
		
	\end{figure}
	
	In Fig.~\ref{fig:DASCH_phased_pdm}, we create a phase plot of the DASCH dataset using the period from our PDM analysis and the time of the secondary's inferior conjunction ($T_0$) from \citet{marsh}. In order to highlight any time-dependent trends in the data, the color of each marker indicates the year of observation. The morphology of the orbital waveform was constant across the observations, and although five measurements from the 1980s appear to be $\sim$0.25~mag brighter than older measurements obtained at similar orbital phases, we demonstrate in Sec.~\ref{sec:checkstars} that this is very likely an instrumental effect. Additionally, while the heterogeneous nature of the DASCH plates induces some scatter in the light curve, the phase of maximum light does not show any obvious drift, as would occur if there were a significant change in the orbital period during the observations. 
	
\begin{figure*}[!ht]
		\begin{center}
		\includegraphics[width=\textwidth]{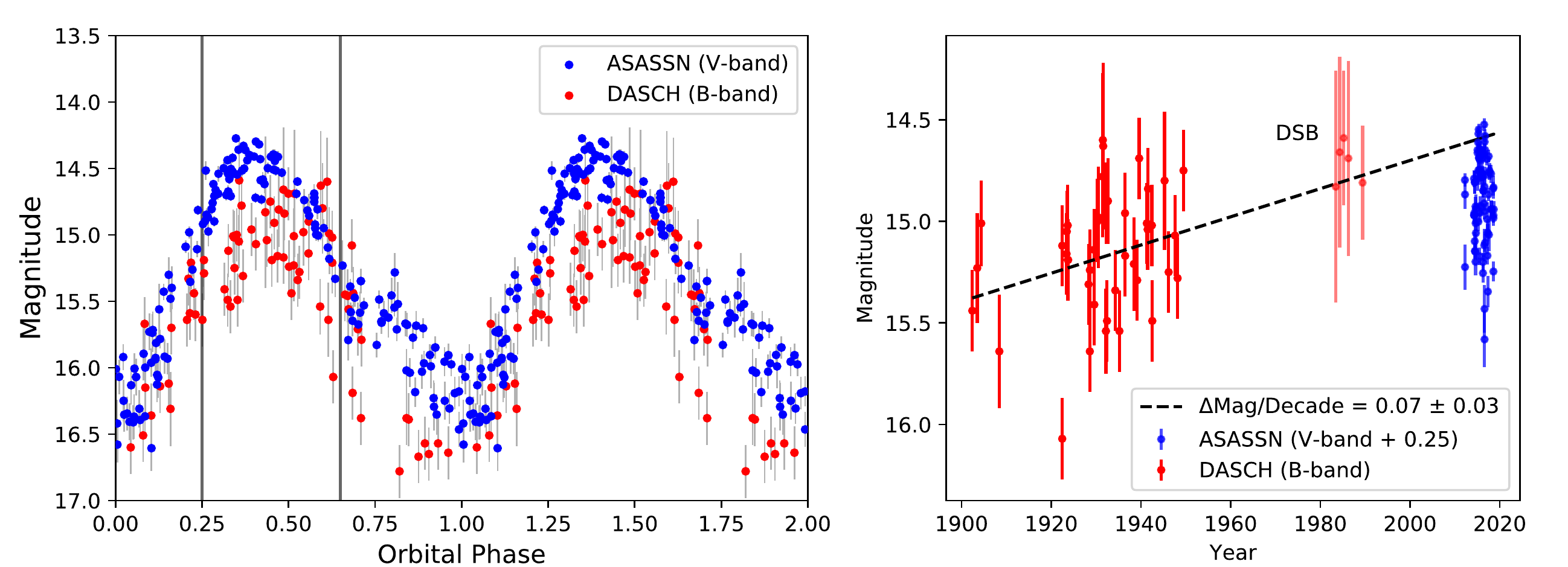}
		\caption{\textbf{Left}: phase plot using our orbital period and $T_{0}$ from \citet{marsh}. Only 20\% of the data points from ASAS-SN were selected to improve visibility. \textbf{Right}: magnitude near orbital maximum, defined as being between orbital phases 0.25 and 0.65, plotted against the year of data collection. The linear fit is to the DASCH data only and is weighted according to magnitude errors, and while it suggests a brightening of 0.07 $\pm$ 0.03 mag/decade, we show in Sec.~\ref{sec:checkstars} and Fig.~\ref{fig:checkstars} that it is likely the result of an instrumental effect in the final five DASCH measurements, each of which is from the Damons South Blue (DSB) series. We also plot the inferred $B$ magnitude of the peak of the orbital waveform in ASAS-SN by adding a $B-V$ of 0.25 \citep{garnavich} to the ASAS-SN $V$ magnitude.}
		\label{fig:light_curve_w/_pdm}
		\end{center}
		
	\end{figure*}

	A comparison of the ASAS-SN and DASCH orbital light curves reveals that the two waveforms are consistent with each other (Fig.~\ref{fig:light_curve_w/_pdm}) with the exception of a systematic brightness offset that is almost certainly the result of a bandpass difference between the $V$-band ASAS-SN photometry and the blue-sensitive DASCH observations. This finding suggests that there was no major change in AR Sco's orbital light curve between the DASCH and ASAS-SN observations. The peak of the orbital waveform is somewhat more clearly defined in the ASAS-SN data, probably because DASCH's long exposures average across much longer segments of the orbital modulation than do ASAS-SN's 90 s exposures.

	\subsection{The apparent long-term brightening is an instrumental effect}
	\label{sec:checkstars}
	
	At first glance, Fig.~\ref{fig:light_curve_w/_pdm} seems to show that AR Sco's peak brightness increased by a quarter magnitude in the 1980s relative to the rest of the DASCH light curve, and a weighted linear fit suggests a gradual increase of 0.07 $\pm$ 0.03 mag/decade. Upon closer examination, however, this rise appears to be an instrumental effect.
	
\begin{figure}
	    \centering
	    \epsscale{1.0}
	    \includegraphics[width=\columnwidth]{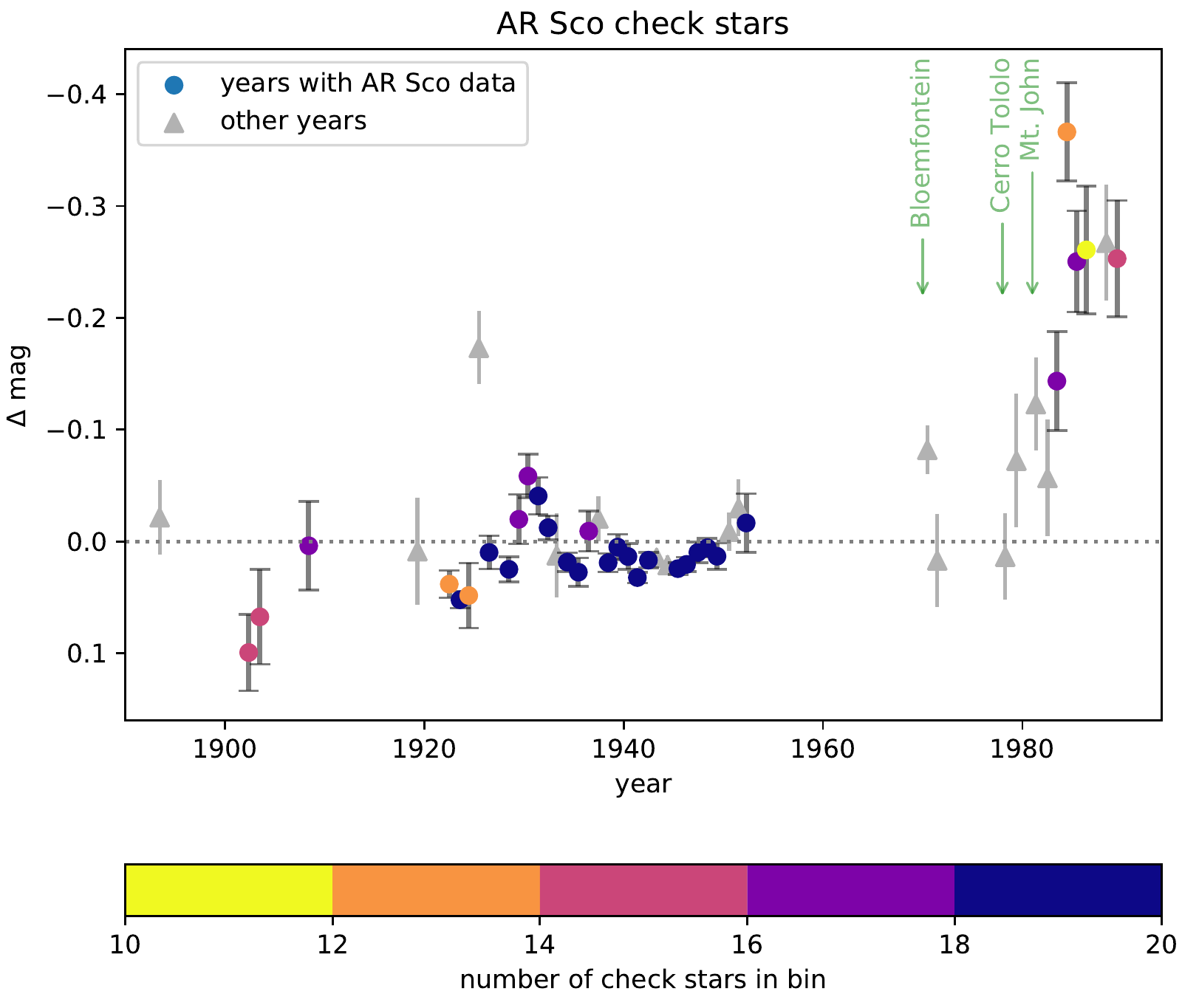}
	    \includegraphics[width=\columnwidth]{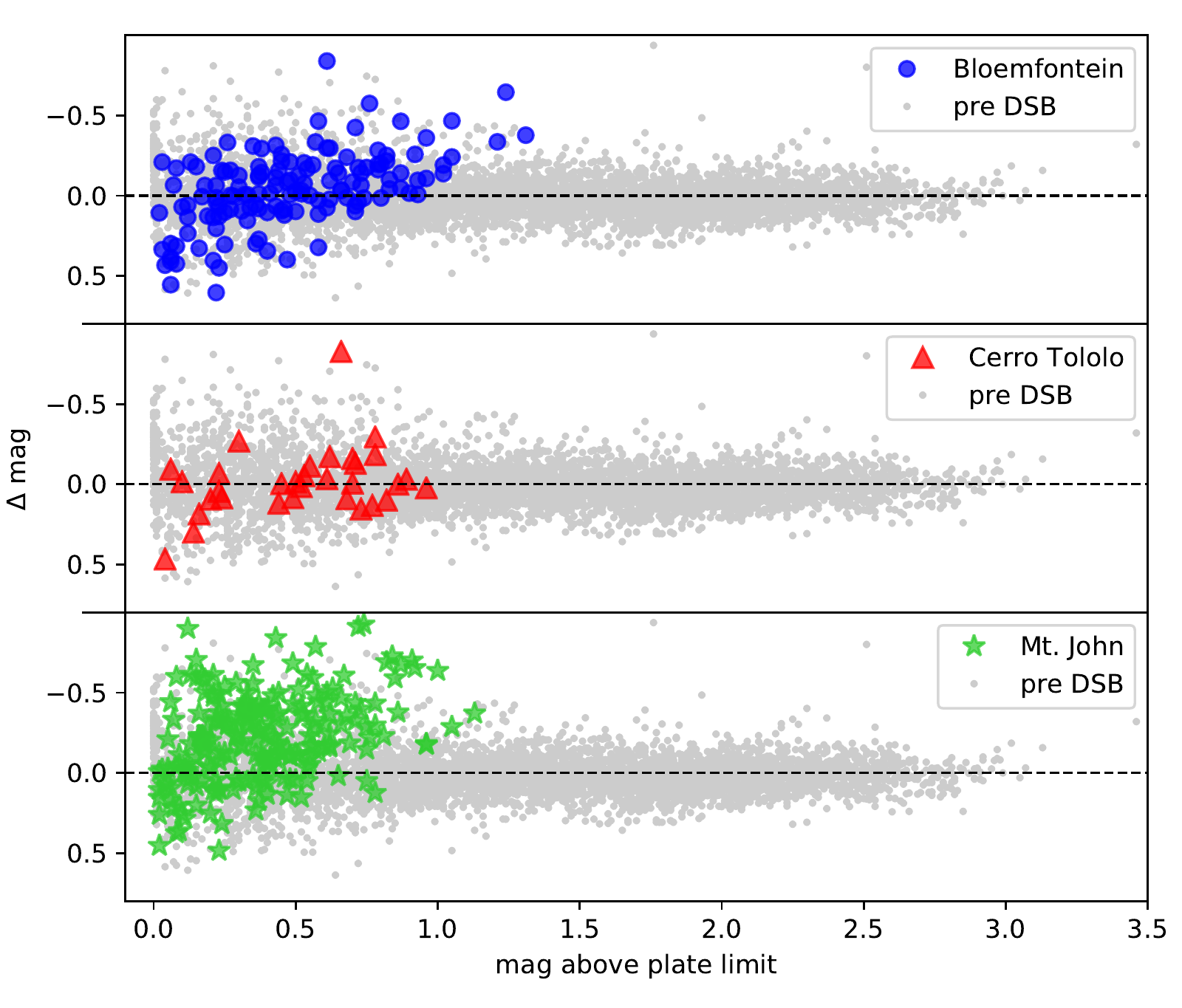}
	    \caption{Analysis of 20 check stars whose colors and brightness are similar to AR Sco. The ordinate, $\Delta$~mag, is the difference between an individual magnitude measurement and that star's mean magnitude in all of DASCH. {\bf Top:} the yearly average $\Delta$~mag for the ensemble of check stars. The intervals for the three DSB observing sites are indicated with vertical arrows. The colored markers, which represent years in which DASCH observations of AR Sco withstood our quality cuts, show a large increase in brightness in the final DSB plates. {\bf Bottom:} behavior of the check stars in the DSB series compared to all other series in DASCH. In the Blomfontein and Mt. John DSB plates, $\Delta$~mag correlates with the proximity of the check stars to the plate's limiting magnitude. Non-DSB data do not show this correlation.}
	    \label{fig:checkstars}
	\end{figure}

	If the rise in AR Sco were of astrophysical origin, it would not be present in check stars analyzed in an identical manner as AR Sco. However, as shown in Fig.~\ref{fig:checkstars}, our ensemble of 20 check stars appeared $\sim$0.25~mag brighter in the final years of DASCH, creating the illusion of a long-term brightening trend similar to the one observed in AR Sco. The effect is even more dramatic when, in an effort to better match the sampling of AR Sco, we limit our analysis of the check stars to years in which there were DASCH observations of AR Sco that withstood our quality cuts.
	
    A close analysis of the DASCH plates reveals a possible systematic issue that might explain this effect. The plates obtained after the Menzel gap in our sample were exclusively from the Damons South Blue (DSB) series, which consisted of plates obtained between 1970-1973 at Bloemfontein, 1978-1979 at Cerro Tololo, and 1981-1989 at Mt. John.\footnote{\url{http://dasch.rc.fas.harvard.edu/HistoryOfPlateSeries/History20.jpg}} In contrast, the pre-Menzel-gap plates came from a variety of sources whose properties differed from the DSB series. In the lower panel of Fig.~\ref{fig:checkstars}, we show that the magnitudes of our check stars in the DSB plates from Bloemfontein and Mt. John plates correlate with how close the stars were to the limiting magnitude of the plate. Stars that were just above the limiting magnitude of the DSB plates tended to have their brightnesses underestimated by the DASCH pipeline, while those that were well above the limiting magnitude were usually brighter than expected. This effect was especially acute in the DSB plates obtained at Mt. John, the exclusive source of the five DSB plates included in our analysis. Fig.~\ref{fig:checkstars} further establishes that the pre-Menzel-gap plates did not show this effect, and it does not appear to be present in the Cerro Tololo DSB plates.


	We stress that our sample of check stars has a very narrow range of properties that approximate the broadband characteristics of AR Sco; it is not a representative sample of all stars observed by DASCH, it would be inappropriate to extrapolate this finding to other DASCH science cases without further analysis. Our work suggests only that near AR Sco's line of sight, relatively blue stars of similar brightness to AR Sco tended to appear anomalously bright in the DSB plates (particularly the Mt. John plates). This tendency refutes the apparent brightening of AR Sco in DASCH. It is worth noting that there has been a recent disagreement between \citet{schaefer} and \citet{hippke} over whether the Damons North Blue series exhibits a flux discontinuity relative to the rest of the DASCH light curve.\footnote{See also Schaefer's response to \citet{hippke} at \url{https://www.centauri-dreams.org/wp-content/HippkeAnalysis.pdf}.} We take no position on their debate. But because the Damons plates are the predominant data source after the Menzel gap, a flux discontinuity---even if it appears only under a very limited set of circumstances---could masquerade as a long-term change in a star's brightness. It would be helpful if a future paper examined this issue in detail.


	\subsection{Frequency Derivative Analysis}
 
	Angular-momentum losses should cause the orbital period to slowly decrease due to a combination of magnetic braking and gravitational radiation. The long baseline of the DASCH observations makes it possible to search for a resulting change in the orbital period. Previously, \citet{stiller} analyzed ASAS-SN observations of AR Sco and used them to constrain any orbital-frequency derivative ($\dot{\nu}$) to be less than $2 \ \times \ 10^{-18}\ \textrm{Hz }  \textrm{s}^{-1} $. 
	
\begin{figure*}[ht!]
	    \centering
	    \includegraphics[width=\textwidth]{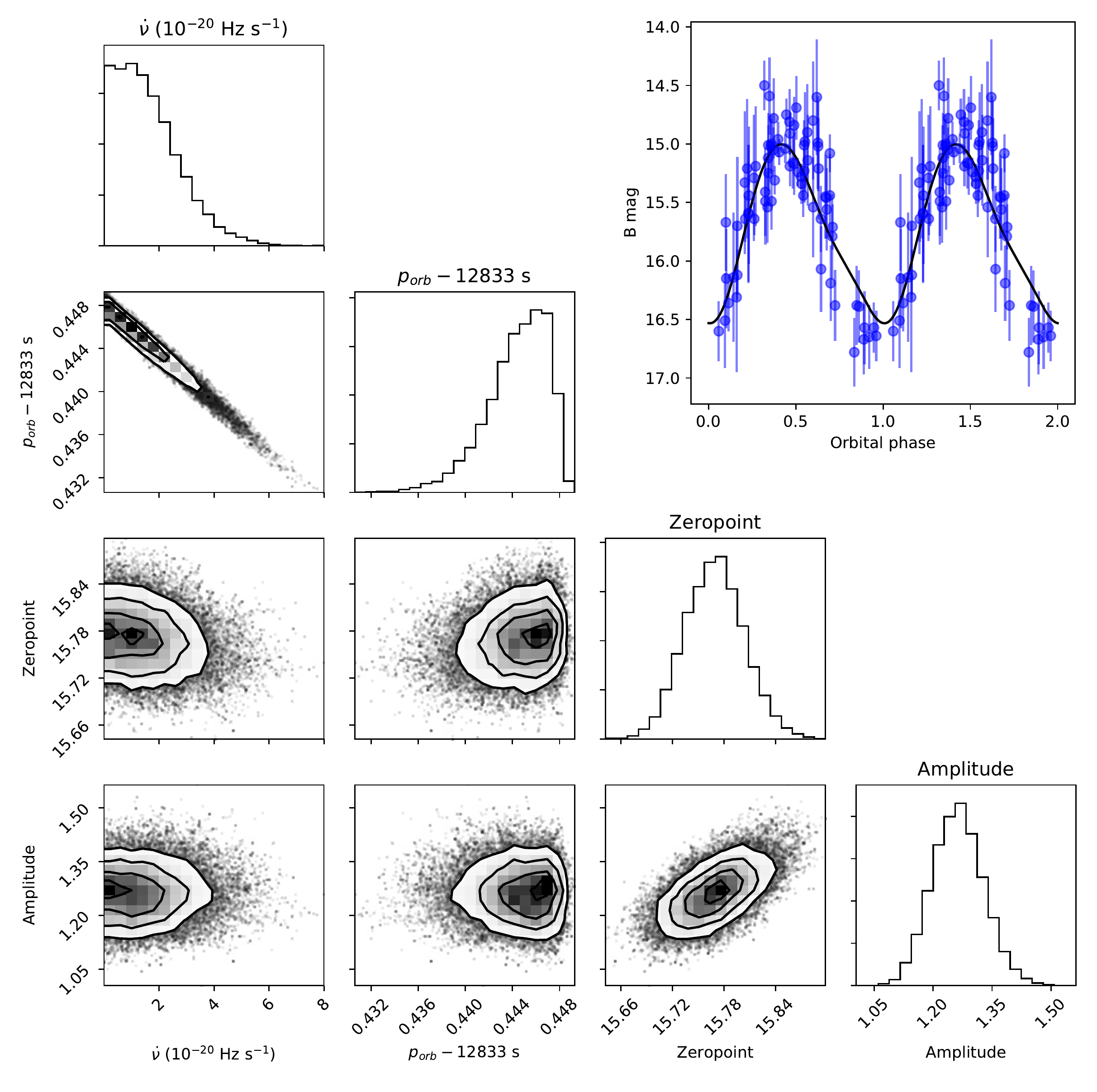}
	    \caption{Corner plot showing the results of the MCMC fitting procedure. The upper right panel shows the DASCH observations phased to a Fourier-series representation of the ASAS-SN light curve (black line). As described in the text, the error bars have been scaled to reflect timing uncertainties. The zero-point is a constant offset added to the ASAS-SN light curve, while the amplitude is a multipicative factor applied to the ASAS-SN data to account for the bandpass difference. The DASCH observations establish a constraint of $\dot{\nu} < 3.8 \times 10^{-20}$ Hz s$^{-1}$ at 95\% confidence.}
	    \label{fig:mcmc}
	\end{figure*}

	We used a Markov chain Monte Carlo (MCMC) procedure to search for evidence of an orbital-frequency derivative. Our	model worked as follows. First, we created a template orbital light curve by taking the ASAS-SN $V$-band photometry, phasing it to the orbital period, and convolving it to match the typical time resolution of the DASCH plates that survived our quality cuts. Next, our MCMC procedure explored different combinations of the orbital frequency and its derivative in an effort to minimize the squared residuals of the DASCH observations from the template ASAS-SN light curve. In addition to the orbital frequency and its derivative, there were two other free parameters: a constant zero-point offset that was added to the template light curve, and a multiplicative factor that scaled the amplitude of the template to match the amplitude of the DASCH observations (Fig. \ref{fig:mcmc}).
	
	Our MCMC model also took into consideration the limited timing accuracy of the plates, which the DASCH project estimates from the number of significant digits from which the times for each plate were recorded. The most common timing uncertainty in the data that we used was $\pm$11 minutes ($\sim$0.05 in orbital phase), a significant source of uncertainty for a fast-varying object like AR Sco. We found that if the timing uncertainties were neglected, the MCMC results suggested the presence of a frequency derivative; however, once we accounted for the timing errors, the posterior distribution for the frequency derivative tailed strongly toward 0, establishing an upper limit on the orbital-frequency derivative of $\dot{\nu} \lesssim3.8\times10^{-20}$ Hz s$^{-1}$. 
	
	To check whether this constraint is reasonable, we plotted the template ASAS-SN light curve and superimposed the phased DASCH observations for various combinations of $\nu$ and $\dot{\nu}$. We found that when $\dot{\nu} \lesssim 5\times10^{-20}$ Hz s$^{-1}$, any value of $\dot{\nu}$ could be offset with a small change to $\nu$, providing confirmation that the DASCH data are not sensitive to such small values of $\dot{\nu}$. However, when we experimented with $\dot{\nu} \gtrsim 5\times10^{-20}$ Hz s${-1}$, it quickly became impossible to find a combination of $\nu$ and $\dot{\nu}$ that caused the DASCH observations to match the template light curve. This by-eye constraint on $\dot{\nu}$ supports the finding of our MCMC analysis that $\dot{\nu} \lesssim3.8\times10^{-20}$ Hz s$^{-1}$.

    Our constraint on the orbital-frequency derivative reinforces the conclusion by \citet{stiller} that the beat-frequency derivative in their dataset is exclusively attributable to a change in the WD spin period and not to the orbital period.

	\section{Discussion}
	
	\subsection{Testing the Precessional Model from \citet{katz}}
	
	If the orbital light curve were the result of uniform irradiation of the companion star's inner hemisphere, the orbital phase of maximum light would coincide with superior conjunction when we view the irradiated inner hemisphere most directly. Instead, as \citet{marsh} first noted, the orbital light curve peaks near $\phi_\textrm{orb}\sim0.4$. As discussed in Sec.~\ref{sec:intro}, \citet{katz} proposed that an obliquity of the WD could produce this behavior and would also result in a precessional period ranging from 20 to 200 yr. \citet{katz} pointed out that an observable consequence of this model is that the phase of orbital maximum would drift on timescales of decades to centuries. 
	
	\citet{katz} and \citet{littlefield} suggested that this hypothesis could be tested by searching for a phase drift in the peak of the orbital maximum. \citet{littlefield} attempted to do so by phasing sparsely sampled, archival light curves to the same photometric ephemeris and searching for evidence of a time-dependent drift of the phase of maximum light. We have determined, however, that their approach does not meaningfully test the precessional model, as the effects of a secular phase drift from precession would precisely mimic an imprecision in the orbital period. If such a phase drift were present in the data, period-finding algorithms would simply report a photometric period that differs very slightly from the true orbital period. To confirm this, we created a synthetic orbital light curve, matched its sampling to the DASCH data, and injected a secular phase drift, similar to that proposed by \citet{katz}. We found that the PDM analysis compensated for the phase drift by reporting a slightly different period than the known period, as expected. Thus, to have had a realistic chance of detecting precession-related phase drift in photometry, \citet{littlefield} would have needed to phase the data to a sufficiently precise spectroscopic measurement of the orbital period. 
	

	The \citet{katz} precessional model can be tested in other ways. For example, it implicitly predicts that the photometric period is the beat period between the precessional period and the orbital period. This inequality of the photometric and orbital periods would be detectable across a sufficiently long spectroscopic baseline. For example, \citet{marsh} measured the time of inferior conjunction of the donor by measuring the radial-velocity variations of its Na~I absorption lines in 2015. Three years later, \citet{garnavich} obtained high time-resolution spectroscopy covering most of an orbit, and the time of inferior conjunction in these spectra can be compared against the \citet{marsh} value to search for any evidence of a difference between the orbital and photometric periods. 
	
	We obtained the \citet{garnavich} spectra and reduced them as described in that paper. We phased the \citet{garnavich} spectra using a linear orbital ephemeris consisting of the  \citet{marsh} time of inferior conjunction and our photometric period, and we then fit a sinusoid to the radial-velocity variations of the K~I $\lambda$7699\AA\ absorption line, which was one of the few absorption features not to be significantly impacted by telluric absorption. The best-fit sinusoid revealed the phase of inferior conjunction to be $\phi = 0.01\pm0.02$. The measurement error suggests that we could have detected a phase shift as small as $\pm0.04$ at $2\sigma$; dividing the three-year spectroscopic baseline by this phase shift suggests that we can rule out any precessional period shorter than $\sim75$~yr at $2\sigma$ confidence. Therefore, there is no evidence of a significant difference between the photometric and spectroscopic periods, and as the spectroscopic baseline increases, the constraints on any WD precessional period will become increasingly rigorous.
	
	Another way of testing the \citet{katz} hypothesis is to search for evidence of an amplitude modulation whose period is consistent with the proposed precessional periods. For instance, \citet{garnavich} showed that if the inner hemisphere of the secondary is uniformly heated and experiences enhanced heating twice each orbit when it passes through the magnetic plane, the observed orbital light curve could be reproduced with remarkable accuracy. The precessional period proposed by \citet{katz} would induce changes in both the amplitude and the shape of the orbital light curve, and the long baseline of the DASCH observations would likely reveal any such variation if it existed. However, in light of the analysis in Sec.~\ref{sec:checkstars}, there is no evidence of such a trend in AR Sco. As \citet{katz} did not predict how the waveform would change across the precessional cycle, it is difficult to use the absence of an amplitude modulation to constrain the length of the precessional period, so our spectroscopic constraint is more quantitatively rigorous.

	We conclude that the precessional model proposed by \citet{katz} is inconsistent with the available data and that the marginal evidence of a time-dependent phase shift noted in Fig.~4 of \citet{littlefield} should not be interpreted as supportive evidence of the precessional model.

	\subsection{Angular-momentum Losses and $\dot{\nu}$}

	The orbital periods of close M-dwarf and WD binaries decrease due to angular-momentum losses, principally from magnetic braking (MB; for systems with orbital periods above $\sim$3~hr) and gravitational radiation (for systems with shorter orbital periods). The observation of orbital decay in some close binaries containing a pair of stellar remnants has provided a triumphant confirmation of the predicted behavior of gravitational radiation, but tests of magnetic braking have proven more difficult for several reasons. For example, there is no universally agreed upon prescription for magnetic braking \citep{knigge}. Moreover, observational measurements of the orbital-period derivative are difficult because they require a long baseline of observations, and other phenomena, such as mass transfer, which can contribute to the period derivative. A system with negligible mass-transfer, like AR Sco, therefore presents an opportunity to measure the orbital decay due to magnetic braking without the complicating effects of mass transfer.

	
	

    

	\citet{knigge} constructed a semi-empirical model of cataclysmic variable (CV) evolution and found that a pairing of scaled versions of angular-momentum losses due to gravitational radiation and the \citet{rappaport} parameterization of MB accurately described the CV evolutionary sequence.  Their revised evolutionary model (Table~4 in \citet{knigge}) predicts that at AR Sco's orbital period, the rate of angular-momentum loss is $\dot{J}_\textrm{MB} = 1.3\times 10^{35}$ dyn cm, compared against an orbital angular momentum of $J = 1.2 \times 10^{51}$ J s. Assuming that $\dot{M}_1 = \dot{M}_2 = 0$, Eq.~9.31 in \citet{warner} yields the corresponding orbital period derivative:

	\begin{equation} 
	\frac{\dot{J}}{J} = \frac{\dot{P}_\textrm{orb}}{3P_\textrm{orb}}.
	\end{equation} 
	After applying the identity
		\begin{equation}\label{eq:nu_dot_to_P_dot}
	    \dot \nu = -\frac{\dot P}{P^2},
	\end{equation}
	the expected $\dot{J}_\textrm{MB}$ from the revised evolutionary track in \citet{knigge} predicts a frequency derivative of $\dot{\nu}_\textrm{MB} \sim 2 \times 10^{-20}$ Hz s$^{-1}$.

	The anticipated $\dot \nu_\textrm{MB}$ is marginally below our upper limit of $\dot{\nu} < 3.8 \times 10^{-20}$ Hz s$^{-1}$, suggesting that over the course of our baseline, the time-averaged rate of angular-momentum loss due to MB could have been at most only a factor of $\sim2$ above the prediction of the \citet{knigge} revised evolutionary model.

	
    
    

    \subsection{Identifying Additional WD pulsars}
    
    \begin{figure}
        \centering
        \includegraphics[width=\columnwidth]{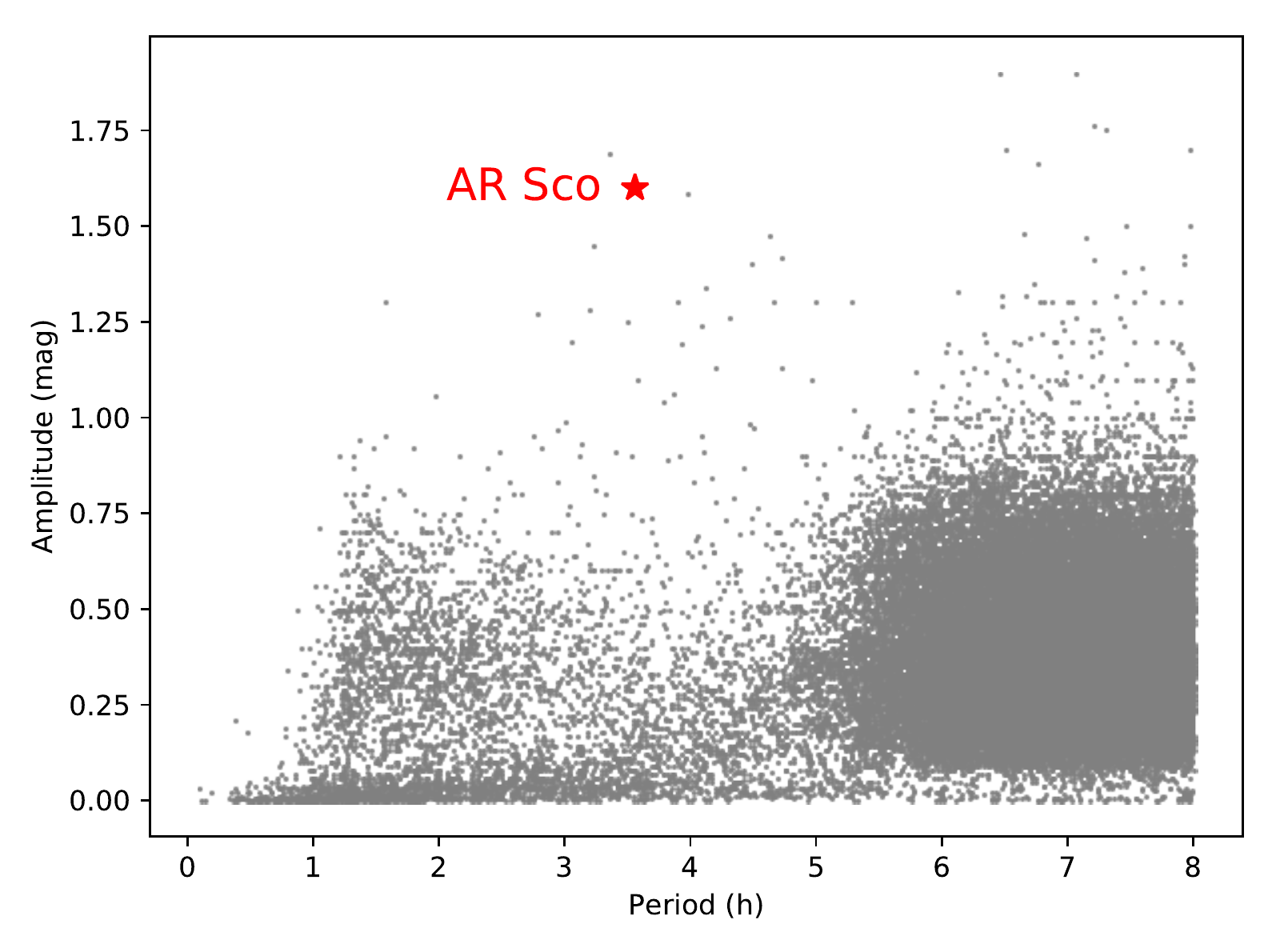}
        \caption{Full amplitude of variability of pulsating stars in the VSX database. We have also included overcontact binaries because of the possibility that some of these stars might be misclassified as pulsating stars if their light curves are poorly sampled; the overcontact binaries account for most of the dense cluster of points above 5~hours. At the exposure times of most surveys, AR Sco-type objects will appear to be highly coherent pulsating stars, but AR Sco's discordantly high amplitude of pulsation suggests a basis for distinguishing WD pulsars from pulsating stars with a similar period. As discussed in the text, this approach is best-suited to WD pulsars with moderate-to-high orbital inclinations.
        \label{fig:amplitude_vs_period}}
    \end{figure}

    AR Sco is the only known WD pulsar and is remarkably close, with a \textit{Gaia} distance of $117.4^{+0.5}_{-0.6}$~pc \citep{bj18}. This proximity suggests that additional WD pulsars might also be relatively nearby and observable, though observational biases have prevented their identification as such. Indeed, as \citet{marsh} noted, AR~Sco was originally classified as a $\delta$~Scuti star, and were it not for fortuitous fast-cadence photometry by amateur astronomers, this system would likely still be consigned to obscurity. The efficient identification of additional AR~Sco-type objects is therefore an important next step in the study of the WD pulsar phenomenon.
    
    By establishing that the orbital light curve of AR Sco has remained remarkably stable for over a century, the DASCH light curve underscores that a WD pulsar's orbital light curve is arguably the most feasible way of identifying candidate WD pulsars in large datasets. While existing surveys generally have insufficient time resolution to detect AR Sco's fast pulsations, they are much more sensitive to high-amplitude, hours-long, and highly coherent periodicities like AR Sco's orbital light curve. The absence of a discernible amplitude modulation of the DASCH light curve means that AR~Sco-type objects would probably appear in these datasets to be extraordinarily high-amplitude $\delta$ Scuti stars, as AR Sco itself was long believed to be. To illustrate how unusual AR Sco's light curve is compared to genuine pulsating stars, we downloaded the periods and amplitudes of all short-period pulsators from the AAVSO VSX database \citep{VSX}, and in Fig.~\ref{fig:amplitude_vs_period}, we indicate how AR Sco's amplitude of orbital variation far exceeds the amplitudes of pulsators with comparable periods.

    A number of all-sky surveys are sensitive to high-amplitude variability with periodicities of a few hours and can be used to identify candidate AR Sco-type objects. Ground-based surveys such as ASAS-SN, ATLAS, and PanSTARRS have the requisite depth and sky coverage to detect candidate AR Sco-type objects. The \textit{Transiting Exoplanet Survey Satellite} (\textit{TESS}) mission is another option, as it is observing a majority of the sky at a 30 minute cadence. However, the low angular resolution of \textit{TESS} means that blending would suppress the apparent amplitude of an AR Sco-type object, so a deblending algorithm would almost certainly be necessary. 
    
    A deep, all-sky catalog of periodic variables that show AR Sco-type waveforms could then be searched for rapid variability to identify true AR Sco-type objects, the successful identification of which would shed light on the evolutionary status of WD pulsars, their occurrence rate, and how they relate with intermediate polars.
    
    An important caveat is that the orbital modulation is undoubtedly sensitive to the orbital inclination, so even if AR Sco's properties are typical of the population of WD pulsars, low-inclination WD pulsars will show much lower-amplitude orbital modulations than AR Sco, whose inclination has been estimated to be $\sim60^{\circ}$ \citep{garnavich}.

    \section{Conclusion}
    
    Using the DASCH light curve to extend the photometric baseline of AR Sco, we establish that the orbital light curve has remained remarkably consistent since the early 20th century. These observations suggest that the mechanism that powers the orbital light curve has remained constant over the course of a century and is a long-term feature of AR Sco.
    
    
    Additionally, we constrain the orbital-frequency derivative to be $\dot{\nu} \lesssim 3.8 \times 10^{-20}$ Hz s$^{-1}$ providing constraints on the efficacy of magnetic braking in this system. Our constraint is in agreement with the predicted orbital-frequency derivative from magnetic braking in the revised evolutionary model in \citet{knigge}, and it reinforces the conclusion from \citet{stiller} that their measurement of the beat-frequency derivative is fully attributable to the spindown of the WD and not to an orbital-frequency derivative.
    

    \acknowledgements
    
    We thank the referee for an expeditious and insightful review and Lindsay Smith for her assistance with analyzing the DASCH data.
    
    The DASCH project is partially supported by NSF grants AST-0407380, AST-0909073, and AST-1313370.


\begin{thebibliography}

\bibitem[Bailer-Jones et al.(2018)]{bj18} Bailer-Jones, C.~A.~L., Rybizki, J., Fouesneau, M., et al.\ 2018, \aj, 156, 58.

\bibitem[Buckley et al.(2017)]{buckley} Buckley, D.~A.~H., Meintjes, P.~J., Potter, S.~B., et al.\ 2017, Nature Astronomy, 1, 29.

\bibitem[Garnavich et al.(2019)]{garnavich} Garnavich, P., Littlefield, C., Kafka, S., et al.\ 2019, \apj, 872, 67.


\bibitem[Grindlay et al.(2009)]{DASCH} Grindlay, J., Tang, S., Simcoe, R., et al.\ 2009, Preserving Astronomy's Photographic Legacy: Current State and the Future of North American Astronomical Plates, ed. W. Osborn \& L. Robbins (San Fransisco, CA: ASP), 101.

\bibitem[Hippke et al.(2016)]{hippke} Hippke, M., Angerhausen, D., Lund, M.~B., et al.\ 2016, \apj, 825, 73.


\bibitem[Katz(2017)]{katz} Katz, J.~I.\ 2017, \apj, 835, 150.

\bibitem[Kochanek et al.(2017)]{kochanek} Kochanek, C.~S., Shappee, B.~J., Stanek, K.~Z., et al.\ 2017, PASP, 129, 104502.

\bibitem[Knigge et al.(2011)]{knigge} Knigge, C., Baraffe, I., \& Patterson, J.\ 2011, The Astrophysical Journal Supplement Series, 194, 28.



\bibitem[Littlefield et al.(2017)]{littlefield} Littlefield, C., Garnavich, P., Kennedy, M., et al.\ 2017, \apjl, 845, L7.

\bibitem[Marsh et al.(2016)]{marsh} Marsh, T.~R., G{\"a}nsicke, B.~T., H{\"u}mmerich, S., et al.\ 2016, \nat, 537, 374.

\bibitem[Potter \& Buckley(2018a)]{pb1} Potter, S.~B., \& Buckley, D.~A.~H.\ 2018, \mnras, 478, L78.

\bibitem[Potter \& Buckley(2018b)]{pb2} Potter, S.~B., \& Buckley, D.~A.~H.\ 2018, \mnras, 481, 2384.

\bibitem[Rappaport et al.(1983)]{rappaport} Rappaport, S., Verbunt, F., \& Joss, P.~C.\ 1983, \apj, 275, 713.


\bibitem[Schaefer(2016)]{schaefer} Schaefer, B.~E.\ 2016, \apjl, 822, L34.

\bibitem[Shappee et al.(2014)]{shappee} Shappee, B.~J., Prieto, J.~L., Grupe, D., et al.\ 2014, \apj, 788, 48.

\bibitem[Stellingwerf(1978)]{PDM} Stellingwerf, R.~F.\ 1978, \apj, 224, 953.

\bibitem[Stiller et al.(2018)]{stiller} Stiller, R.~A., Littlefield, C., Garnavich, P., et al.\ 2018, \aj, 156, 150.


\bibitem[Warner(1995)]{warner} Warner, B. 1995, Cataclysmic Variable Stars (Cambridge: Cambridge Univ. Press).

\bibitem[Watson et al.(2006)]{VSX} Watson, C.~L., Henden, A.~A., \& Price, A.\ 2006, Society for Astronomical Sciences Annual Symposium, 25, 47.

\end{thebibliography}
\end{document}